\documentclass[11pt,a4paper]{article}

\usepackage{graphicx}
\usepackage{setspace}

\def\ltsima{$\; \buildrel < \over \sim \;$}

\def\lsim{\lower.5ex\hbox{\ltsima}}
\def\gtsima{$\; \buildrel > \over \sim \;$}
\def\gsim{\lower.5ex\hbox{\gtsima}}

\begin{document}

\title{\bf A distortion of very--high--redshift galaxy number counts by gravitational lensing}

\author{J. Stuart B. Wyithe$^1$, Haojing Yan$^2$,
Rogier A. Windhorst$^3$, Shude Mao$^{4,5}$}

\date{}

\maketitle

\begin{small}
\noindent 
$^1$School of Physics, University of Melbourne, Parkville, Victoria 3010, Australia\\
$^2$Center for Cosmology and AstroParticle Physics, The Ohio State
University, Columbus, OH 43210\\
$^3$School of Earth and Space Exploration, Arizona State
University, Tempe, AZ 85287-1404\\
$^4$Jodrell Bank Centre for Astrophysics, University of
Manchester, Manchester, M13 9PL, UK\\
$^5$National Astronomical Observatories of China, Chinese Academy of Sciences, Beijing 100012, China\\
\end{small}

\noindent {\bf The observed number counts of high-redshift galaxy candidates$^{1-8}$
have been used to build up a statistical description of star-forming
activity at redshift $z\gsim7$, when galaxies reionized theUniverse$^{1,2,9,10}$.
Standard models$^{11}$ predict that a high incidence of gravitational
lensing will probably distort measurements of flux and number of
these earliest galaxies. The raw probability of this happening has
been estimated to be $\sim0.5$ per cent (refs 11, 12), but can be larger
owing to observational biases. Here we report that gravitational
lensing is likely to dominate the observed properties of galaxies with
redshifts of $z\gsim12$, when the instrumental limiting magnitude is
expected to be brighter than the characteristic magnitude of the
galaxy sample. The number counts could be modified by an order
of magnitude, with most galaxies being part of multiply imaged
systems, located less than 1 arcsec from brighter foreground
galaxies at $z\approx2$. This lens-induced association of high-redshift
and foreground galaxies has perhaps already been observed among
a sample of galaxy candidates identified at $z\approx10.6$. Future surveys
will need to be designed to account for a significant gravitational
lensing bias in high-redshift galaxy samples.
}

Along random lines-of-sight, the {\em raw} probability (or optical depth)
for multiple imaging of objects at high redshifts --- owing to gravitational lensing by individual foreground field galaxies$^{11,12}$ --- is $\simeq0.5\%$. However, all galaxy populations are observed to have a characteristic luminosity ($L_\star$), brighter than which galaxy numbers drop exponentially, and below which numbers rise with a very steep power-law slope$^{1,4,6}$.  The potential for gravitational lensing to modify the observed statistics therefore increases dramatically, owing to the magnification of numerous, intrinsically faint galaxies to observed fluxes that are above the survey limit. This effect, which is known as {\em magnification bias}$^{13}$,
leads to an excess of gravitationally lensed galaxies among flux-limited samples.
Magnification bias is expected to be particularly significant at high redshifts ($z\gsim8$), where current observations may only be probing the exponential tail of the LF$^{4}$, so that the number density could be rising very rapidly towards the detection limit. Indeed, multiply imaged candidates at $z\gsim 7$ have already been discovered behind
foreground clusters via targeted searches$^{14-16}$,
demonstrating this to be an efficient method for finding
faint high redshift galaxies$^{14,17}$.

We assess magnification bias among high redshift galaxies assuming singular, spherical, isothermal gravitational lenses, which produce one or two images, and designate the apparent magnitude of the more magnified image (the only image in the absence
of lensing) as $m_{\rm AB,1}$. We then calculate, as a function of the assumed 
characteristic luminosity (expressed in terms of absolute magnitude $M_\star$), the fraction of galaxies brighter than the magnitude limit 
$(m_{\rm lim})$ for the Hubble Ultra Deep Field ({\em HUDF}) that would be multiply imaged (designated $F_{\rm lens}$). Such curves are shown at 
$z = 6,$ 7, 8.6 and $10.6$ in panel {\bf a} of Figure~1. The superimposed solid and open
points correspond to lens fractions for different
estimates$^{1,4}$ of $M_\star$ at these redshifts. At
$z\simeq6-7$, we expect only $\simeq1$ percent of galaxies to be lensed. At
$z\simeq8-10$, however, we expect a lensed fraction of a few to a few
tens of percent, depending on the true value of $M_\star$. Note that since current survey
limits are significantly fainter than $M_\star$ at $z\simeq6-7$, the lens
fraction is quite insensitive to $M_\star$. However, at higher redshifts where the
survey limits might be much closer to $M_{\star}$, the lensing fraction is very
sensitive to its uncertain value. 

Predictions for a significant lens fraction at $z\gsim8$ stand
in apparent contrast to the fact that no image pairs have been identified in
the {\em HUDF}. However, we find the probability that a multiply imaged galaxy, with 
observed $m_{\rm AB,1}$ has a corresponding {\it second} image with 
$m_{\rm AB,2}<m_{\rm lim}$ (i.e. detectable with the {\em HUDF} data) to be only
$\simeq10\%$, even for galaxies that are one magnitude brighter than
$m_{\rm lim}$ (see Supplementary Information). Thus, as shown in panel {\bf b} of Figure~1, the fraction of galaxies $(F_{\rm mult})$ that are detected as 
multiply imaged systems in the {\em HUDF} is an order of magnitude lower than the true lensed fraction. Although this fraction would increase somewhat if elliptical lenses were included in our analysis,  multiply imaged systems are not expected to be observed in the current data. On the other hand, magnification bias also leads to a concentration of high redshift sources --- both singly and multiply imaged --- around foreground galaxies$^{18-20}$. The resulting correlation between high redshift candidates and bright foreground galaxies therefore offers an alternative avenue to observing the effect of gravitational lensing. A schematic diagram illustrating this point, as well as magnification bias, is included as Supplementary Figure~1. 

To quantify this correlation, we first determine the distribution of separations between random lines-of-sight and the nearest bright ($H\leq25$ mag) foreground 
galaxy in the {\em HUDF}, measured as the angular distance to the centroid. This is shown by the dotted black line in panel {\bf c} of Figure~1.  This distribution can be compared to the predictions of our model 
(dashed line in panel {\bf c} of Figure~1). If the candidate sample consists of both multiply imaged {\it and} unmagnified galaxies, then
the observed distribution of separations should be
a weighted sum of the random and the lensed line-of-sight distributions. The
correct weighting is the probability for gravitational lensing, $F_{\rm lens}$.
Two examples are shown in panel {\bf c} of Figure~1. 
The fraction of galaxies found within $\Delta \theta\simeq1-2$ arcseconds of a foreground galaxy is very sensitive to the characteristic luminosity if $M_\star\gsim-19$ mag, providing a potential observable for the influence of lensing on the number counts of $z\gsim8$ candidates.

For comparison with the lensing predictions, we 
have measured the distribution of separations between a sample of $z\sim10.6$ candidates$^{4}$ and their nearest bright ($H\leq25$ mag) foreground galaxy.
Comparing the distributions, we find that these candidates are observed to be closer to 
bright foreground galaxies than are random lines-of-sight. On the other hand,  
the candidates are found at larger separations from foreground galaxies than 
would be predicted if they were all multiply imaged. 
Quantitatively, the Kolmogorov-Smirnov probabilities between 
the observed distributions and the {\it all-random} model or the 
{\it all-lensed} model (see Supplementary Information) indicate that both models are
rejected at high significance. This suggests that a 
fraction of candidates may be gravitationally lensed. Moreover, we have generated the distribution of redshifts for foreground 
galaxies found within $\Delta\theta<1.5$ arcseconds of the $z\sim10.6$
candidates. These distributions are consistent with the distribution of gravitational lens
redshifts, while the redshift distribution of all bright foreground galaxies are not, which supports the hypothesis that foreground galaxies are lensing a fraction of the $z\sim10.6$ candidates into the observed sample.

With the introduction of the James Webb Space Telescope ({\em JWST}), galaxy surveys will be undertaken out
to even higher redshifts, well into the epoch of First Light$^{21}$. Panels {\bf a} and {\bf b} of Figure~2 show $F_{\rm lens}$ as a function of
$M_\star$ out to $z=20$. The flux limits correspond to
an ultra-deep survey ($m_{\rm lim}=31.4$~mag), and a medium-deep survey
($m_{\rm lim}=29.4$~mag).  The evolution of the characteristic luminosity is unknown at these unexplored redshifts.
 For comparison, we therefore plot squares corresponding
to estimates of $M_\star$ based on an extrapolation from lower redshift {\em HUDF} data$^{1}$. 
Figure~2 shows that in ultra-deep {\em JWST} surveys for First Light objects at $z\gsim$14,
more than $F_{\rm lens}\sim10\%$ of the candidates could be lensed. In much shallower {\em JWST} surveys
that only sample the exponential tail of the Schechter LF, a lensed
object fraction of $F_{\rm lens}\sim10\%$ could be seen at redshifts as low as $z\sim8$--10. 
However at $z\gsim$14, the lensed
fraction in such surveys could be much higher, and may even represent the 
majority of observed galaxies. Surveys with {\em JWST} will therefore need
to be carefully planned and analyzed
to account for the influence of foreground lensing galaxies. 

As in the case of the {\em HUDF}
the fraction of galaxies that will be detected as 
multiply imaged systems by {\em JWST} is significantly lower than the true multiple image fraction. However, as the multiple image fraction becomes very large at high redshifts, observed doubles could become common;
larger than $F_{\rm mult}\sim10\%$ at redshifts $z\gsim12$ in a medium-deep ($m_{\rm AB}<29.4$~mag) {\em JWST} survey, and $z\gsim16$ in an ultra-deep ($m_{\rm AB}<31.4$~mag) survey. Panels~{\bf e} and {\bf f} present the predicted distributions of separation for galaxies discovered by {\em JWST} from bright foreground galaxies. If the observed evolution in $M_\star$ continues to higher redshift, then the spatial distribution of high redshift galaxies relative to foreground galaxies will depart from random at redshifts $z\gsim14$ for ultra-deep surveys, and at $z\gsim10$ for medium-deep surveys with {\em JWST}. A crucial prediction is that the majority of very high redshift galaxies discovered with {\em JWST} may be located less than 1 arc-second from a bright foreground galaxy, and will have been gravitationally magnified into the sample.

A key goal for {\em JWST} will be to measure the number counts of high redshift candidates, and to construct luminosity functions (LF)
in order to build up a statistical description of star-forming activity in
galaxies. LFs describing the density of sources per unit luminosity are parametrised by a Schechter function$^{22}$,
$\Psi(L) \propto\left({L}/{L_\star}\right)^\alpha\exp{(-{L}/{L_\star})}{1}/{L_\star}$,
including free parameters for the power-law
slope at low luminosities ($\alpha$), and the characteristic absolute AB-magnitude 
[$M_{\rm AB}-M_\star=-2.5\log_{10}(L/L_\star)$] brighter than which galaxy numbers drop exponentially. Importantly, 
gravitational lensing has the potential to significantly modify the observed LF from its intrinsic shape$^{23}$. 
In particular, at very high luminosities in the exponential tail of the Schechter function, the LF shape can be modified from exponential to power-law, since gravitational lensing magnifies numerous faint sources to apparently higher luminosities.
Figure~3 shows that the shapes of LFs near the flux limit are not affected by gravitational lensing at $z\approx6-8$. However, {\em if} the evolution of the galaxy LF continues into the reionisation era (we assume an extrapolation of the fitting formulae based on candidates discovered in and around the {\em HUDF}\,$^{1}$), then we find that {\em JWST} will measure LFs that are significantly modified by lensing at redshifts above $z\sim14$ and $z\sim10$ in its ultra-deep and medium-deep surveys, respectively.

Our results imply that 
while published LF's at $z\gsim7$ are not currently corrected for a potential gravitational lensing bias, such corrections will need to be prescribed in detail for future surveys that aim to
measure the build-up of stellar mass among the first
galaxies using {\em JWST}. In particular, studies of the high redshift LF will require good understanding of the magnification bias for high redshift galaxies, in order to correct for gravitational lensing and uncover its true unlensed shape at $z\gsim12$. Of particular importance will be the unknown
evolution of $M_\star$, which could be influenced (for example) by
supernovae feedback from population-III stars$^{24}$, in addition to
hierarchical clustering and formation. 
Gravitational lensing could magnify $z\gsim$10--12 objects to flux levels that will allow spectroscopic observations using JWST and the
largest ground-based near-IR spectrographs. 
A further implication of our analysis is that gravitational lensing could be used
to probe the shape of the high redshift LF at luminosities that are not
otherwise accessible$^{12}$, using the association of high redshift
galaxy candidates and foreground galaxies, combined with careful modelling of the gravitational lensing bias.

\small
\singlespace
\bigskip
\bigskip
\bigskip
\bigskip
\bigskip
\bigskip
\bigskip
\bigskip
\bigskip
\bigskip
\bigskip
\bigskip
\noindent
{\bf Acknowledgments:}\\ The authors thank K.-H. Chae. JSBW was supported in
part by a QE-II fellowship and grants from the Australian Research Council.
HY acknowledges supports of the long-term fellowship program of the Center for
Cosmology and AstroParticle Physics (CCAPP) at The Ohio State University. HY and RAW were supported by grants from the Space Telescope Science Institute, which is
operated by the Association of Universities for Research in Astronomy, Inc. RAW was supported by a NASA JWST Interdisciplinary Scientist grant.

\normalsize
\vskip 0.2in

\bigskip
\bigskip
\bigskip
\bigskip
\bigskip

\newpage

\singlespace

\begin{figure*}
\vspace{-10mm}
\begin{center}
\includegraphics[width=17cm]{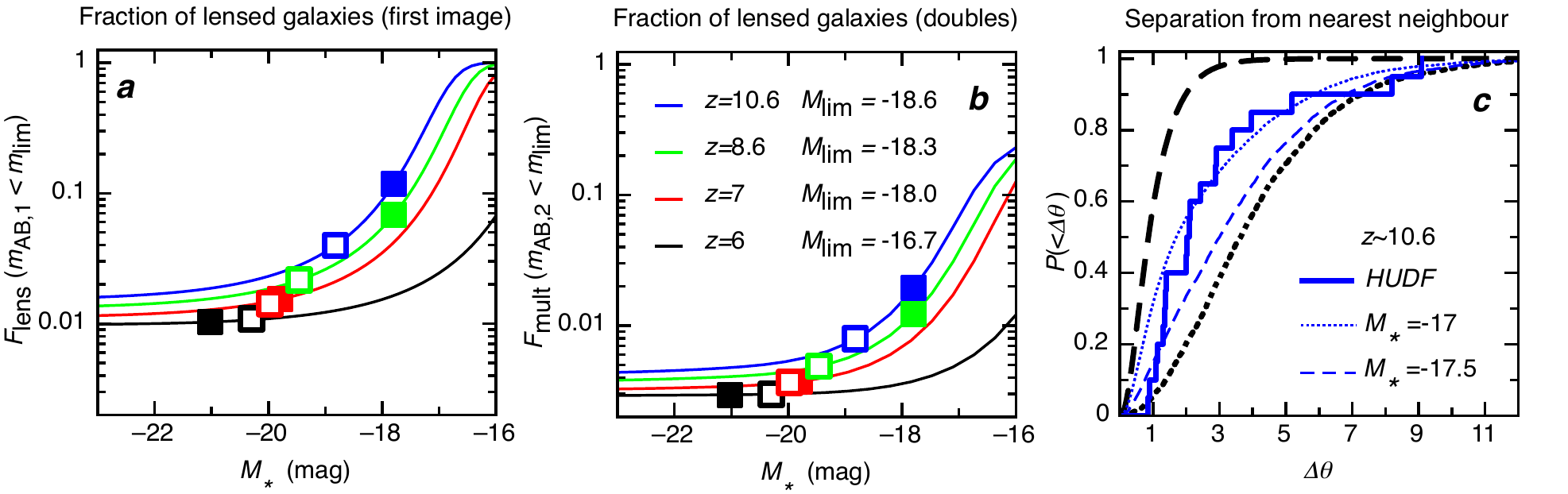}
\end{center}
\vspace{-0mm}
\begin{small}
\noindent Figure 1. {\bf Gravitational lens fractions among candidate high redshift {\em HUDF} galaxies.} 
{\bf Panel a:} The fraction of multiply-imaged high redshift galaxies.
{\bf Panel b:} The fraction of high redshift galaxies 
in which multiple images could be detected in the {\em HUDF}.
{\bf Panel c:} The probability distribution of image separations (at $z\simeq10.6$) relative to the nearest bright foreground galaxy, in the cases of random lines-of-sight (black dotted line), of gravitational lenses (black dashed line), and for composite distributions computed for two (faint) values of $M_\star$. Also shown is the distribution of measured separations for twenty $z\sim10.6$ candidates$^{4}$ in the {\em HUDF} (stepped blue histogram). Lyman-break
galaxy candidates have been selected with median redshifts of $z\simeq6$, 
$z\simeq7$, $z\simeq8.6$ and $z\simeq10.6$.
At $z\simeq6$, candidate selection using the Advanced Camera for Surveys
reaches$^{6}$ $m_{\rm lim}=30$ mag (absolute magnitude $M_{\rm
lim}=-16.7$ mag). At higher redshifts, objects in the {\em WFC3} {\em HUDF} data can be
selected$^{4}$ to $m_{\rm lim}\simeq 29.0$~mag, corresponding to $M_{\rm lim}=-18.0$, $-18.3$ and $-18.6$~mag at $z\simeq7$, 8.6 and 10.6. 
The open squares correspond to lens fractions given
the fitting formula$^{1,25}$ $M_\star\simeq -21 + 0.32\times(z-3.8)$. The solid squares
represent alternative estimates$^{4,6}$ of $M_\star$. The model for gravitational lensing$^{26}$ is based on the velocity dispersion function
of galaxies$^{27}$. Galaxy mass distributions are modelled
as Singular Isothermal Spheres, and we assume a constant co-moving density 
of lenses. Elliptical lenses would not significantly alter the cross-section$^{28}$, but would provide additional images, and so increase the fraction of observed galaxies that are lensed.
We assume a
Schechter LF$^{22}$, with power-law slope$^{1}$ $\alpha=-2$. A change of 0.3 in $\alpha$ leads to a 40\% change in the lens probability. 
We have used the cosmology based on 7-year results from the {\em WMAP}
satellite$^{29}$ throughout this {\em Letter}. 
\end{small}
\end{figure*}

\clearpage
\begin{figure*}
\begin{center}
\includegraphics[width=17cm]{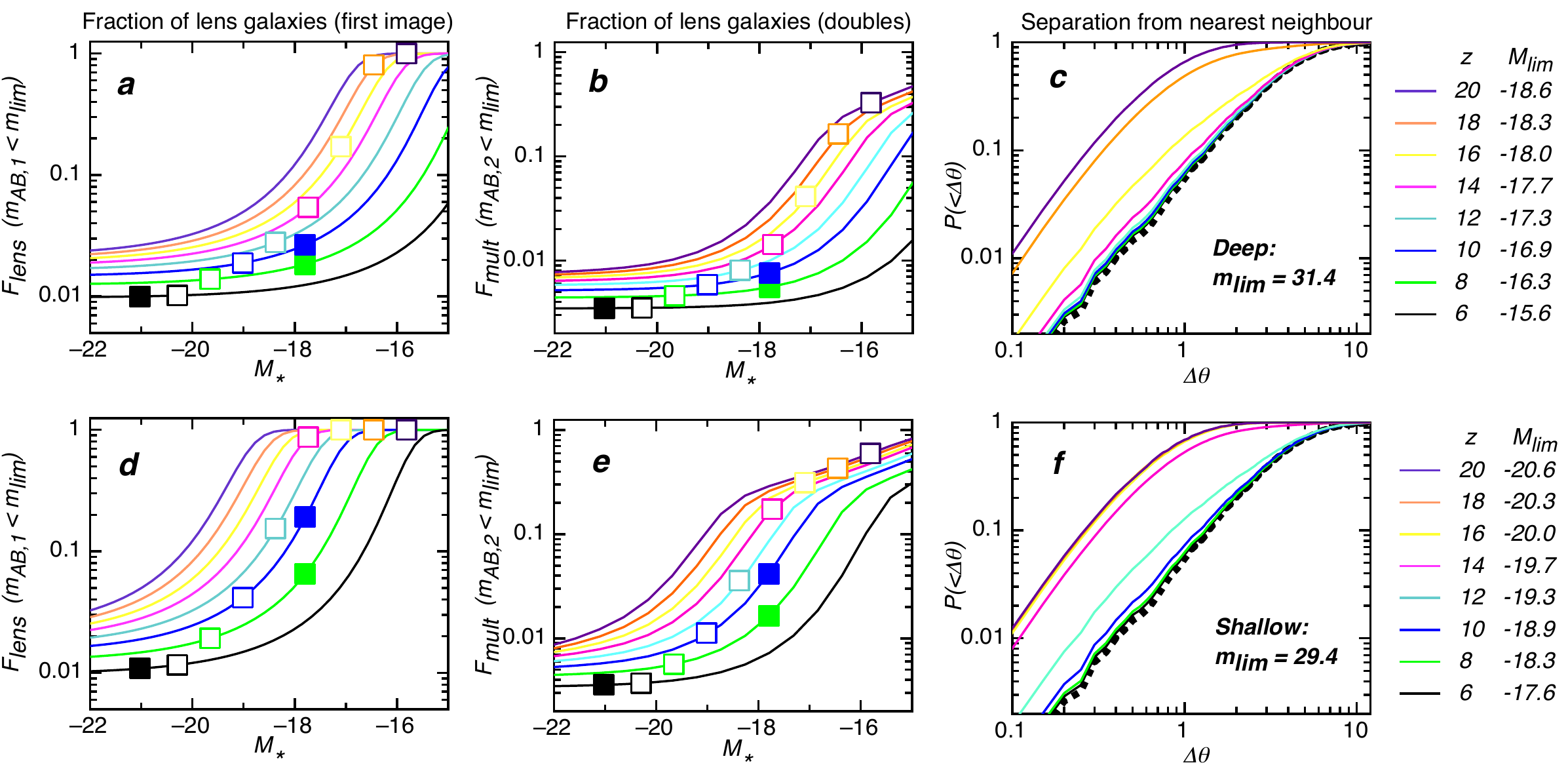}
\end{center}
\begin{small}
\noindent Figure 2. {\bf Probabilities for multiple imaging of high redshift
galaxies to be observed with {\em JWST}}. The panels mirror those of Figure~1, but with examples of limiting magnitudes and redshifts  appropriate for both an {\em ultra-deep} survey ($m_{\rm AB}<31.4$ mag, $\sim 1$\,nJy), and a {\em medium-deep} survey ($m_{\rm AB}<29.4$ mag) with {\em JWST}. The
corresponding limiting absolute magnitudes are listed. {\bf Panels a-b:} The fraction of
observed galaxies that have multiple images. The
superimposed solid and open points correspond to lens fractions given a
faint value$^{4}$ of $M_\star=-17.8$ at $z\sim8.6$ and $z\sim10.6$, and a fitting formula $M_{\star}(z)$ based on lower redshift data, respectively$^{1,25}$. The latter is extrapolated to high redshift where data does not yet exist. {\bf Panels c-d:} The fraction of high redshift galaxies 
in which multiple images could be detected by {\em JWST}. {\bf Panels e-f:} The probability distribution of image separations relative to the nearest bright foreground galaxy, in the cases of random lines-of-sight (black dotted line), and for composite distributions computed for values of $M_\star$ extrapolated from observations in the {\em HUDF} using the previously mentioned fitting formula$^{1,25}$. 
We note that imaging surveys with {\em JWST} will be working at the diffraction limit ($\sim0.08$ arcseconds resolution FWHM) at $\sim2\,\mu$m. This resolution is higher than is currently available in the {\em HUDF} near-IR images, where candidates have been selected in close proximity to bright foreground galaxies, and hence high redshift candidates will also  be detectable close to foreground galaxies.
\end{small}
\end{figure*}

\clearpage
\begin{figure*}
\begin{center}
\includegraphics[width=10cm]{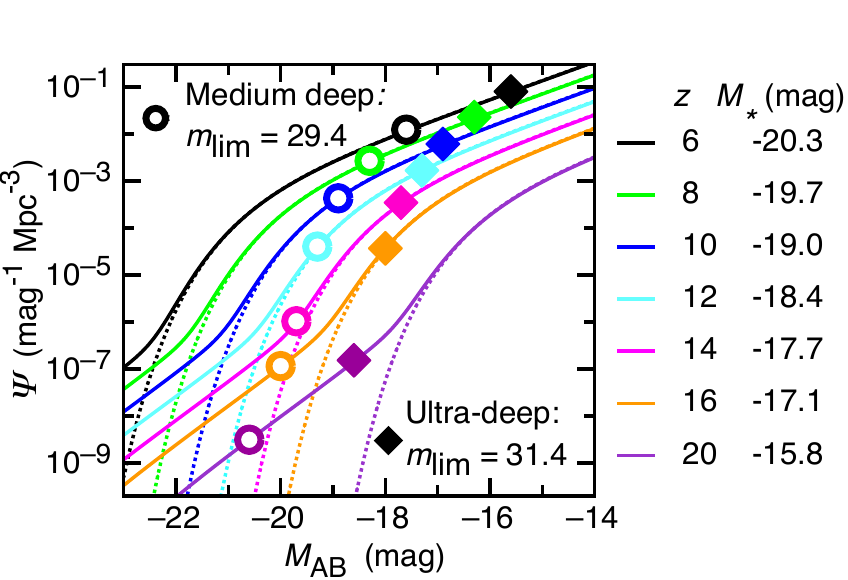}
\end{center}

\begin{small}
\noindent Figure 3. {\bf Gravitational lens induced modification of the bright end of the high redshift
galaxy luminosity function to be observed with {\em JWST}.} Thin curves present the intrinsic LF ($\Psi$), and solid curves the observed LF following modification from gravitational lensing. For simplicity, a uniform magnification was assumed outside regions of sky that are multiply-imaged, with a value such that flux is conserved over the whole sky. The parameters describing the LF are extrapolated to high redshift, where data does not yet exist, assuming fitting formulae based on data from the {\em HUDF}\,$^{1,25}$. Of particular relevance are the values of $M_\star$, which are listed. The solid and open points show the luminosities and densities of the faintest galaxies to be observed with {\em JWST}, assuming limiting magnitudes appropriate for both an {\em ultra-deep} {\em JWST} survey ($m_{\rm AB}<31.4$ mag), and a {\em medium-deep} {\em JWST} survey ($m_{\rm AB}<29.4$ mag). The probability for gravitational lensing will become of order unity in the steep exponential parts of the LF at sufficiently high redshifts. This 
{\em gravitational forest} should not to be confused with the purely mathematical 
effects of image crowding that 
makes the detection and de-blending of faint objects harder at progressively
fainter fluxes$^{30}$. These latter effects are referred
to as either the {\em instrumental confusion limit} --- when the instrumental
resolution is not good enough to statistically distinguish {\it all}  faint
background objects from brighter foreground objects --- or the {\em natural
confusion limit} --- when the instrumental resolution is good enough to
distinguish faint background  objects from brighter foreground objects, but the
images are so deep that  objects start overlapping because of their own
intrinsic sizes. The {\em HUDF} and {\em JWST} images are in the latter regime$^{30}$, and as argued in this {\em Letter},
likely have the {\it additional} fundamental limitation that {\it gravitational
lensing} will magnify a non-negligible fraction of faint objects into
the sample.
\end{small}
\end{figure*}

\clearpage

\begin{center}
\begin{Large}
\title{---\\{\it SUPPLEMENTARY INFORMATION}\\---}
\end{Large}
\end{center}

\singlespace

\noindent The {\em Letter to Nature} estimates the probability for gravitational lensing among high redshift galaxies, with emphasis on current surveys using the {\em HUDF}, and future surveys to be carried out using {\em JWST}. The {\em Letter} also  uses the observed distributions
of separation between very high redshift galaxy candidates in the {\em HUDF} and foreground galaxies, to show that a significant  fraction
of these objects are likely to be gravitationally lensed. The
following sections expand the brief descriptions of the modelling and interpretation that can be found
in the {\em Letter to Nature}.

\section{Schematic picture of magnification bias and foreground galaxy
correlation}

In Supplementary Figure~1, we present a schematic representation of a portion of the
{\em HUDF}, which shows how magnification bias leads to a correlation between
foreground galaxies and high redshift candidates. Panel {\bf a} shows a
representation of the Schechter function$^{22}$, which describes
the luminosity function (LF) of high redshift galaxies. The limiting absolute magnitude
$M_{\rm lim}$ and characteristic magnitude $M_\star$ are shown for reference.
Gravitational lensing magnifies sources relative to their intrinsic luminosity,
and draws intrinsically faint galaxies into the flux limited sample. Since
faint galaxies are much more common than bright galaxies, the number of sources
per unit area in regions of lensing magnification is significantly higher. This
leads to a bias of sources near foreground galaxies. To illustrate this effect
on the high redshift galaxy samples in the {\em HUDF}, we sketch in panel {\bf b} a
portion of the sky approximately 10 arcseconds across. In this panel,
background sources (i.e. high redshift galaxies) are shown in red and
foreground galaxies (those near $z\simeq 1-2$) in blue. The faint galaxies (with
$M_{AB}>M_{\rm lim}$) are signified by open symbols, while the closed symbols
signify bright galaxies with $M_{AB}<M_{\rm lim}$. The black dotted disks
denote regions of sky where background sources will be multiply-imaged by a 
foreground galaxy. For illustration, this schematic representation 
overestimates the total lensing cross-section, which is $\simeq0.5\%$, by a factor
of $\simeq10$. The typical angular scale of these regions is 1 arc-second. We show
those faint galaxies that lie within these lensing regions in green. In panel
{\bf c,} the faint galaxies that are close enough to bright foreground galaxies
to be multiply-imaged (shown in green), producing in general
a bright image with $M_{AB}<M_{\rm lim}$, and an undetected faint image with
$M_{AB}>M_{\rm lim}$. Finally, the observed association of high redshift
galaxies with bright foreground galaxies --- once gravitational lensing bias
has been accounted for --- is shown in panel {\bf d}. In this example 2 of the
5 observed high redshift galaxies ($M_{AB}<M_{\rm lim}$) have entered the
sample owing to gravitational magnification, and have close alignment with
foreground galaxies as a result. In this case we find 40\% of high redshift galaxies within
$\simeq1$ arc-second of bright foreground galaxies, even though the observed
density of bright foreground objects is 1 per 20 square arcseconds. We note that gravitational lensing can also lower the observed density of sources on the sky that have neighbouring foreground galaxies by magnifying the angular
extent of the image plane relative to the source plane. This effect, which is
usually referred to as {\em depletion}, is not dominant when the LF is steep, as is the case for high redshift galaxies.

\section{Lens model}

We refer to the a-priori probability for a galaxy at redshift $z_{\rm gal}$ to be
multiply-imaged by an intervening foreground galaxy as the multiple image
optical depth$^{17}$ 

\begin{equation}
\tau_{\rm m}= \int_{0}^{z_{\rm gal}} \frac{d\tau_{\rm m}}{dz} dz, 
\end{equation}

\noindent where

\begin{equation}
\frac{d\tau_{\rm m}}{dz} = \int d\sigma \Phi(\sigma,z) (1+z)^3 \frac{cdt}{dz} \pi D_{\rm d}^2 \theta_{\rm ER}^2(\sigma,z), 
\end{equation}

\noindent $\theta_{\rm ER}$ is the Einstein radius as a function of velocity dispersion
$\sigma$ and redshift $z$, $D_{\rm d}$ is the angular diameter distance to
the lens, and $t$ is time. To calculate $\tau_{\rm m}$, we use the expression for the angular
Einstein radius for a Singular Isothermal Sphere (SIS)

\begin{equation}
\theta_{\rm ER}(\sigma,z)=0.9''\,\,\frac{D_{\rm ds}}{D_{\rm s}}\left(\frac{\sigma}{161\mbox{km/s}}\right)^2,
\end{equation}

\noindent where $D_{\rm s}$ and $D_{\rm ds}$ are the angular diameter distances to the
source, and between the lens and source, respectively. 

To evaluate $\Phi(\sigma,z)$, we first assume$^{26}$ the {\em Sloan
Digital Sky Survey} (SDSS) velocity dispersion function$^{27}$
$\Phi_{\rm SDSS}(\sigma)$

\begin{equation}
\Phi_{\rm SDSS}(\sigma) d\sigma = \Phi_\star\left(\frac{\sigma}{\sigma_\star}\right)\frac{\exp{[-(\sigma/\sigma_\star)^\beta]}}{\Gamma(\alpha/\beta)}\beta\frac{d\sigma}{\sigma},
\end{equation}

\noindent where $\Phi_\star=2\times10^{-3}$Mpc$^{-3}$, $\alpha=2.32$, $\beta=2.67$ and
$\sigma_\star=161\,$km/s. We further assume that the lens population has a constant co-moving density $\Phi(\sigma,z) = \Phi_{\rm SDSS}(\sigma)$. Although the 
density of galaxies must decline at high redshift, this approximation is reasonable, since most lensing occurs at 
$z\lsim1.5$. The uncertainty in predictions of the lens fraction owing to the unknown evolution of the velocity 
dispersion function is approximately a factor of two$^{26,31}$\,.
We note that this 
prescription gives a lensing cross-section for $z\simeq 2$ quasars that is 
consistent with the SDSS analysis$^{26}$, which is an observational
requirement. The lens model assumes that galaxy velocity dispersions reach down to as low as
$\sigma=10$km/s. However, as the lensing neighbours are selected by velocity
dispersion, the distribution of lensed separations is not sensitive to the
assumed cutoff, because the lens cross-section is proportional to velocity dispersion to the fourth power ($\sigma^4$).

We have utilised a simple lens model. In particular we have not included non-spherical lens distributions, which produce four rather than two image lenses in some cases. Indeed, empirical estimates for the fraction of quasar lenses that have four images of about 40\% have been obtained from the homogeneous CLASS sample ({\tt http://www.aoc.nrao.edu/$\sim$smyers/class.html}), and of about 15\% from the Sloan Digital Sky Survey$^{32}$\,. While the predicted four-image to two-image ratio depends on the ellipticity of the lensing galaxies$^{32}$\,, the ellipticity is found not to significantly influence the overall cross-section for multiple imaging$^{28,33,34}$. On the other hand, the magnification bias can be larger for an elliptical lens, which would increase the expected multiple imaging rate$^{28}$. Moreover, the additional images in a four image lens would increase the fraction of observed candidates that are part of a multiply-imaged galaxy. Using spherical lenses for our estimates is therefore conservative with respect the expected influence of gravitational lensing on samples of high redshift galaxy candidates, both in terms of the number of lenses predicted and the association between high redshift candidates and bright foreground galaxies. We note here that the lens population for $z\sim8-10$ candidates is at higher redshift than the lens galaxies responsible for the aforementioned samples. However the measured ellipticity distribution is nearly constant over a very wide range of flux and redshift$^{35}$\,.
Thus, we argue that since our simple model provides a good statistical description of the available data, neglecting elliptical lenses is reasonable, particularly given the range of other uncertainties.

\subsection{magnification bias}

Flux limited samples are subject to magnification bias, which increases the
relative probability that detected galaxies are gravitationally
lensed$^{17}$, and concentrates sources in a flux limited sample
around foreground objects$^{18}$. Yan et al.$^{4}$ have
observed a number of $z\simeq 8-10$ candidates that have neighbouring
bright foreground galaxies. As discussed in the {\em Letter}, this correlation  is likely
to be the manifestation of these effects. The magnification bias for sources with
observed luminosities between $L$ and $L+dL$ is

\begin{equation}
B(L) = \frac{\int_{\mu_{\rm min}}^{\mu_{\rm max}}\frac{d\mu}{\mu} \frac{dP}{d\mu} \Psi(L/\mu)}{\Psi(L)}{},
\end{equation}

\noindent while the corresponding overall magnification bias in a flux limited sample is

\begin{equation}
B_{\rm lens} = \frac{\int_{\mu_{\rm min}}^{\mu_{\rm max}}d\mu \int_{L_{\rm lim}}^\infty dL\frac{dP}{d\mu} \Psi(L/\mu)}{\int_{L_{\rm lim}}^\infty dL\frac{dP}{d\mu} \Psi(L)},
\end{equation}

\noindent where $dP/d\mu$ is the probability distribution for magnification ($\mu$) within the
range $\mu_{\rm min}<\mu<\mu_{\rm max}$. Of relevance for high redshift surveys
in the  {\em HUDF} or with  {\em JWST} (which have an angular resolution much better than the image separation$^{30}$)
is the magnification distribution for the brighter image

\begin{equation}
\frac{dP_{{\rm m},1}}{d\mu} = \frac{2}{(\mu-1)^3} \hspace{5mm}\mbox{for}\hspace{3mm} 2<\mu<\infty.
\end{equation}

\noindent We adopt a Schechter$^{22}$ function for the LF

\begin{equation}
\Psi(L) dL = \Psi_\star\left(\frac{L}{L_\star}\right)^\alpha\exp{(-\frac{L}{L_\star})}\frac{dL}{L_\star},
\end{equation}

\noindent where $\Psi_\star$ is the characteristic density in Mpc$^{-3}$, and $\alpha$ is
the power-law slope at luminosities below the characteristic break at
$L_\star$. Below, and in the {\em Letter}, we quote the characteristic
luminosity in terms of the absolute magnitude
$M_\star=M+2.5\log_{10}{L/L_\star}$. 

\subsection{gravitationally lensed luminosity function}

We note that in the presence of significant gravitational lensing, the LF can be modified from its intrinsic form$^{23}$, leading to a power-law slope at the bright-end of $-3$ (as shown in Figure~3 of the {\em Letter}).
The modified LF can be estimated by modelling the overall magnification distribution using the probability distribution for magnification of multiply-imaged sources over a fraction $\tau_{\rm m}$ of the sky, combined with a de-magnification $\mu_{\rm demag}=(1-\langle\mu_{\rm mult}\rangle\tau_{\rm m})/(1-\tau_{\rm m})$ elsewhere. Here $\langle\mu_{\rm mult}\rangle=4$ is the mean magnification of multiply-imaged sources, and $\mu_{\rm demag}$ has been calculated in order to conserve flux on the cosmic sphere centred on an observer. The modified LF can then be approximated using the expression
\begin{equation}
\Psi_{\rm obs}(L) = (1-\tau_{\rm m})\frac{1}{\mu_{\rm demag}}\Psi(L/\mu_{\rm demag}) +  \tau_{\rm m} \int_0^\infty d\mu \frac{1}{\mu} \left(\frac{dP_{\rm m,1}}{d\mu} + \frac{dP_{\rm m,2}}{d\mu} \right) \Psi(L/\mu),
\end{equation}
where $dP_{\rm m,2}/d\mu=2/(\mu+1)^3 $ for $0<\mu<\infty$, is the probability distribution for the second image. We approximate the true magnification distribution by using a constant value of $\mu_{\rm demag}$ in regions of no multiple imaging. This is valid for the modification of the LF at luminosities much brighter than $M_\star$, in which we are interested in this work.

\section{Lensing predictions for high redshift surveys}

We summarise the predictions of our lensing model in Supplementary Figure~2. As shown
in panel {\bf a}, the lensing optical depth rises toward high
redshift$^{11}$, and is 4-5 times as large for sources at $z\simeq6$
as at $z\simeq1.5$. It doubles again from $z=6$ to $z=20$, so that at $z>10$ the
multiple imaging fraction is greater than 0.5\%, even in the absence of
magnification bias. Panel {\bf b} shows the magnification bias as a function
of the difference between $M_\star$ and the survey limit in absolute magnitude
$M_{\rm lim}$. At low redshifts, deep surveys can probe well below $M_\star$,
so that the magnification bias is dominated by the power-law slope ($\alpha$)
of the Schechter function at low luminosities, and the resulting bias is of
order unity. At very high redshifts, however, current surveys can only reach
$M_\star$ or even brighter, and hence the bias can be much higher (tens or
hundreds) owing to the exponential nature of the LF sampled. We next
combine the optical depth $\tau_{\rm m}$ with the bias $B_{\rm lens,1}$ to find the
multiple image fraction $F_{\rm lens}=B_{\rm lens,1}\tau_{\rm m}/(B_{\rm
lens,1}\tau_{\rm m}+(1-\tau_{\rm m}))$, where we have assumed the bias of those galaxies which
are not multiply-imaged to be unity. In panel {\bf c} we plot contours of
$F_{\rm lens}$ as a function of $z$ and ($M_\star-M_{\rm lim}$). Surveys at low
redshift ($z\lsim3$), with limits fainter than $M_\star$, should have multiple image
fractions below 1\%. However, at higher redshifts the lens fraction can be much
higher. For example, a survey at $z\gsim6$ that reaches only 1 magnitude brighter
than $M_\star$ could have a lens fraction of 10\%. Current and
future surveys at $z>6$ with {\em HST} and {\em JWST} lie in this upper-right portion of
panel {\bf c}. Only ultradeep surveys with {\em JWST} that reach well below $M_\star$
at $z\gsim$10 will have their lensing fraction drop well below 10\% again.

\section{High redshift galaxy candidate samples}

To compare the predictions of our model with samples of high redshift galaxy candidates, we investigate samples from the {\em HUDF} compiled by Yan et al.$^{4}$. These and other authors  have employed the Lyman-break
(or {\em dropout}) technique to select galaxies at $z\gsim 7$ in the {\em HUDF}.
We note that the major colour criteria used to select the samples of Yan et al.$^{4}$ are very
similar to those employed by other groups including Bouwens et al.$^{1,3}$. However the overlap of individual candidates among the samples from
these two teams is small. In particular, none of $J$-dropouts compiled by Yan et
al.$^{4}$ are
among the three $J$-dropouts presented by Bouwens et al.$^{1}$.
There could be a range of reasons for this disjoint. With respect to our
current work, we note that one reason for the difference in sample selection
could be the choice of whether to include candidates near bright foreground objects.
By construction, the samples of Yan et al.$^{4}$ were not biased
against regions around foreground objects, indicating that {\it if} gravitationally lensed, multiply-imaged galaxies do exist in the {\em HUDF} at $z\sim8-10$, then they would be selected. We therefore concentrate here on the
predicted gravitational lensing statistics for these samples.

The $z\approx 8.6$ sample used to discuss the gravitational lensing of galaxies in the {\em HUDF} as part of this work consists of 15 $Y$-dropouts (spanning the redshift range of $7.7\lsim z\lsim 9.4$), while the $z\approx 10.6$
sample consists of 20 $J$-dropouts (spanning $9.4\lsim z\lsim 11.8$).
These objects are all very faint, and have magnitudes ranging from $M_{\rm AB}= 28.0-29.0.$

\subsection{lensing predictions for $z\simeq8-10$ candidates}

We have 
calculated multiple-imaging probabilities for the $z\simeq8-10$ samples$^4$ as a function of galaxy
absolute magnitude assuming $M_\star=-17.8$ mag. These results  can be used to discuss
lensing probabilities for individual $z\simeq8-10$ dropout
candidates$^{4}$ in more detail. 

Panel {\bf a} of Supplementary Figure~3
shows the probability that a galaxy with absolute magnitude $M_{AB,1}$ is multiply-imaged.
At $z\simeq6-7$, only galaxies much brighter than $M_{AB,1}<-21$ mag have a
significant chance of being lensed. However, at $z\simeq8-10$ galaxies as faint as
$M_{AB,1}\simeq-19$ mag have a substantial lens fraction. 
Of course, these are
just statements reflecting the relative brightness of $M_{\rm lim}$ and $M_*$. 
Our results suggest that a number of $z\simeq8-10$ galaxies detected in the {\em HUDF}
should be multiply-imaged. On the other hand, we note that we have not
identified any image pairs in the {\em HUDF}. 
Panel {\bf b} shows the probability that a {\em lensed} galaxy with observed
$M_{AB,1}$ has a corresponding {\it second} image with $M_{AB,2}<M_{\rm lim}$, 
such that it is also detectable above the {\em HUDF} flux limit. For this probability
to be large ($\gsim50\%$), the detected image must be more than $\simeq 1$~mag
brighter than $M_{\rm lim}$. Panel {\bf c} shows the fraction of galaxies that
are part of a lensed pair in which {\it both} images are detectable, [$F_{\rm
dbl}=F_{\rm lens} \times P(M_{AB,2}<M_{\rm lim}|M_{AB,1})$]. 
We find that at $z\simeq6-7$, only galaxies that are several magnitudes brighter than $M_{\rm
lim}$ have a reasonable chance (few-10\%) of being observed as a multiple
image system. However, at $z\simeq8-10$, this probability increases to $\gsim10\%$
for galaxies that are only a magnitude brighter than $M_{\rm lim}$.

We note that the predicted fraction would increase if we modelled
elliptical lenses which can have more than two  images. We roughly
estimate the fraction in this case by noting that a four-image lens
typically has either two bright images of approximately equal
magnification where the source is near a fold caustic, or three
bright images with the central one having a magnification equal to
the sum of the other two$^{36}$\,. Thus, close to the detection
limit, we expect either the two bright images, or only the brightest
of three bright images would be detected for typical four-image
lenses. We therefore argue that for the (empirically observed)
15-40\% of cases where the lens has four images, the fraction of
multiply-imaged systems in which more than one image is detected will
increase by at most a factor of approximately two.

In Supplementary Figure~3, we have superimposed squares to show probabilities for
individual galaxy candidates in the {\em HUDF} $^{4}$\,. 
We use $M_\star=-17.8$~mag estimated by Yan et al.$^{4}$ as an example. By summing probabilities for
individual galaxy candidates in the Yan et al.$^{4}$ sample, we 
calculate the (mean) expected number of lensed systems, finding
$\langle N_{\rm lens}\rangle=0.8\pm0.1$ and 
$\langle N_{\rm lens}\rangle=1.7\pm0.2$ among the 15 and 20 candidates at 
$z\simeq8.6$ and $z\simeq10.6$, respectively. If the true $M_\star$ value is
fainter, these numbers will be higher. A Poisson distribution with mean 
$\langle N_{\rm lens}\rangle = 2.5$ implies that at least one lens pair would
be found among the observed $z\simeq8-10$ sample in 92\% of cases, which stands
in apparent contrast to the fact that no image pairs have been identified in
the {\em HUDF}. However, we find the probability that a {\em lensed} galaxy with 
observed $m_{AB,1}$ has a corresponding {\it second} image with 
$m_{AB,2}<m_{\rm lim}$ (i.e. detectable with the {\em HUDF} data) to be only
$\simeq10\%$, even for galaxies that are one magnitude brighter than
$M_{\rm lim}$. Here we neglect the caveat that secondary images could fall on
top of the foreground galaxies, 
which would further reduce the chance of their being observed. We estimate that the number of systems that would be
observed as doubles (i.e. both images detected) to be 
$\langle N_{\rm dbl}\rangle=0.2\pm0.06$ and 
$\langle N_{\rm dbl}\rangle =0.4\pm0.1$ at $z=8.6$ and $10.6$, respectively. A
Poisson distribution with mean $\langle N_{\rm dbl}\rangle = 0.6$ implies 
that the observed $z\simeq8-10$ sample would not contain any doubles in most
(55\%) cases. Thus, with $M_\star=-17.8$, we find that $N_{\rm lens}\simeq2-3$
of the detected galaxies in each redshift range should be multiply-imaged, but
do not necessarily expect any of these to be identified as multiple image
systems. On the other hand, an even fainter value of $M_\star=-17.3$ ($-17.1$)
implies that at least one double would be observed in 90\% (99\%) of cases,
imposing an upper-limit of $M_\star\lsim-17$ at $z\sim8-10$. We note that the values of $M_\star$ --- as measured --- could also 
be biased by gravitational lensing (see Figure 3 of the {\em Letter}). Currently, none of the published LFs at $z\gsim7$ are corrected for the potential lensing bias.
However it is clear from the results presented in our {\em Letter}, that such corrections will need to be prescribed in detail in the future.

The mean magnification of detected lensed images (with $M_\star=-17.8$) is
$\langle \mu\rangle\simeq6$, indicating that gravitational lensing in these
samples would lead to over-estimates of the luminosity density at $z\simeq8.6$
and $z\simeq10.6$ of $\simeq50\%$ and $\simeq80\%$, respectively, if the
magnification is neglected. 
Since the axis ratio of lensed images is equal to the magnification for an SIS,
this also implies that the lensed images should be significantly elongated, and
indeed some candidates appear to have this property$^{4}$\,. However,
the signal-to-noise for the detected candidates is too low to draw quantitative
conclusions.

\section{Distribution of lensed separations}

As shown in the previous section, we find that in most cases only the more magnified image will be brighter
than the detection threshold. We therefore calculate the expected distribution of angular separation between a
lens galaxy and the brighter of the two images. The apparent angular separation of the bright image with magnification $\mu$ from the center of a lensing SIS at redshift $z<z_{\rm gal}$ is

\begin{equation}
\Delta\theta_{\rm lens}(\mu,z) = \left(1+\frac{1}{\mu-1}\right)\theta_{\rm ER}(z).
\end{equation}

\noindent Using this expression we evaluate the probability distribution for the
separation of bright images of image pairs from the lensing galaxy 

\begin{equation}
\frac{dP}{d\Delta\theta} \propto \int_0^{z_{\rm gal}}dz \int_{2}^{\infty}d\mu\int_{L_{\rm lim}}^\infty dL \frac{d\tau_{\rm m}}{dz}\frac{dP_{\rm m,1}}{d\mu}\Psi(L/\mu)\delta_{\rm dir}([\Delta\theta-\Delta\theta_{\rm lens}(\mu,z)],
\end{equation}

\noindent where $\delta_{\rm dir}$ is the Dirac delta function, and $L_{\rm lim}$ is the unlensed luminosity corresponding to the survey flux limit.

\section{Observed correlation between high redshift candidates and foreground galaxies}

For comparison with the lensing predictions, we 
measure the 
distribution of separations between $z\simeq8-10$ candidates$^{4}$ and their nearest bright ($H\leq25$ mag) foreground galaxy.
The red histograms in panels {\bf a} of 
Supplementary Figures~4 and 5 show the cumulative distributions of this separation for the 
$z\simeq8.6$ and $z\simeq10.6$ candidates, respectively. 
Comparing the distributions in these two panels with the random line-of-sight and lensed predictions, two trends are obvious. Firstly, these $z\simeq8-10$ candidates are observed to be closer to 
bright foreground galaxies than are random lines-of-sight. On the other hand,  
the candidates are found at larger separations from foreground galaxies than 
would be predicted if they were all multiply-imaged. 
Quantitatively, the Kolmogorov-Smirnov probabilities ($P_{\rm KS}$) between 
the observed distributions and the {\it all-random} model or the 
{\it all-lensed} model (labeled in the figure) indicate that either model is
rejected at high significance. This suggests that a 
fraction of these candidates may be gravitationally lensed.

For illustration, the thick black
lines in panels {\bf a} of Supplementary Figures~4 and 5 show the composite
distributions corresponding to multiple image lens fractions of
$F_{\rm lens}=0.2$ and 0.4, at $z\simeq8.6$ and 10.6, respectively. These provide an
excellent fit to the data ($P_{\rm KS}$ values labeled in Supplementary Figures~4 and 5). 
Panels {\bf b} of Supplementary Figures~4 and 5 show the differential distributions
of the observed angular separations (red), as well as the corresponding 
composite models (thick black) and the random distributions (dotted black). The
latter demonstrates that the largest observed separations can be attributed to
a random distribution. 

  We next examine the redshift distributions of the nearest bright foreground
galaxies, using spectrophotometric redshift estimates$^{37}$\,. The cumulative distributions are compared in 
panels {\bf c}  of Supplementary Figures~4 and 5 for the $z\simeq 8.6$ and 10.6
candidates, respectively. The red histograms are the distributions for the
neighbours of the high redshift candidates, the dotted black lines are distributions for the
neighbours of random lines-of-sight, and the dashed black lines are distributions for the 
expected gravitational lens redshift$^{17}$.
The redshift distributions of the
foreground galaxies associated with the full samples of $z\simeq8-10$ candidates cannot be
differentiated from those associated with random lines-of-sight. In addition,
for the $z\simeq 10.6$ case in particular, foreground galaxy redshifts 
are found not to be drawn from a lensed galaxy population. 
However, the lens angular separation cuts off sharply
at $\Delta\theta\simeq1.5$ arcseconds.
We therefore generate the distribution of redshifts only for foreground 
galaxies found within $\Delta\theta<1.5$ arcseconds of the $z\simeq8-10$
candidates, which are shown as the blue histograms in these two panels.
These distributions are consistent with the distribution of gravitational lens
redshifts, which supports the hypothesis that many close candidate--foreground
galaxy pairs in this sample result from magnification bias. 
In panels {\bf c} and {\bf d} of Supplementary Figures~4 and 5, we show the model redshift distributions corresponding to the values
$F_{\rm lens}=0.2$ and 0.4 for candidates at $z\simeq 8.6$ and 10.6,
respectively (thick black lines). These again provide an excellent fit to the data, which,
when taken together with the correlation between high redshift and foreground galaxy positions, provides compelling evidence for a significant lens fraction among the $z\gsim8$ galaxy candidates, since these foreground galaxies were selected only on the basis of their alignment with high redshift candidates.

\newpage

\begin{figure*}[t]
\begin{center}
\includegraphics[width=15cm]{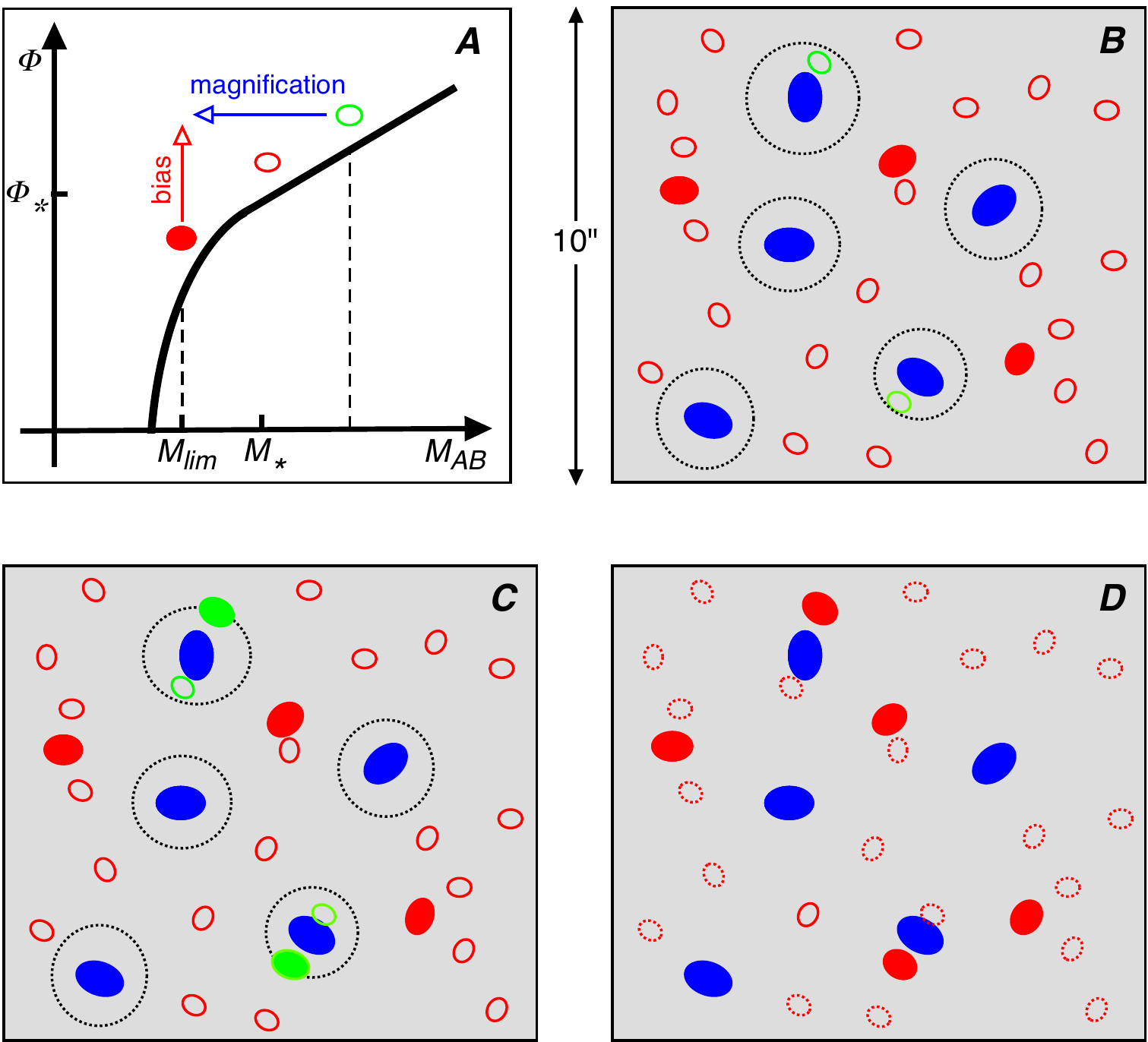}
\end{center}
Supplementary Figure 1. {\bf Schematic representation showing how magnification bias
leads to an association between foreground galaxies and high redshift
candidates.} {\bf Panel a:} The Schechter LF of high redshift galaxies. {\bf
Panel b:} High redshift galaxies (red) and foreground galaxies (blue). Faint
galaxies (those with $M_{AB}>M_{\rm lim}$) are signified by open symbols, while
the closed symbols signify bright galaxies with $M_{AB}<M_{\rm lim}$. The black
dotted disks denote regions of sky where background sources will be multiply
imaged by the foreground galaxy. Faint background galaxies that lie within these lensing
regions are shown in green. {\bf Panel c:} The lensed faint galaxies
are multiply-imaged, producing a bright image with $M_{AB}<M_{\rm lim}$, and an
undetected faint image with $M_{AB}>M_{\rm lim}$. Galaxies located near the lines of sight to foreground galaxies that are not multiply imaged, are deflected to larger separations, resulting in a lowering of observed source density (an effect known as depletion). {\bf Panel d:} The
correlation of observed high redshift galaxies (solid red symbols) with bright
foreground galaxies once gravitational lensing bias has been accounted for. The depletion effect is opposite in sign to the correlation introduced through strong lensing, but is sub-dominant in the case of high redshift galaxies. 
\end{figure*}

\clearpage

\begin{figure*}[t]
\begin{center}
\includegraphics[width=17cm]{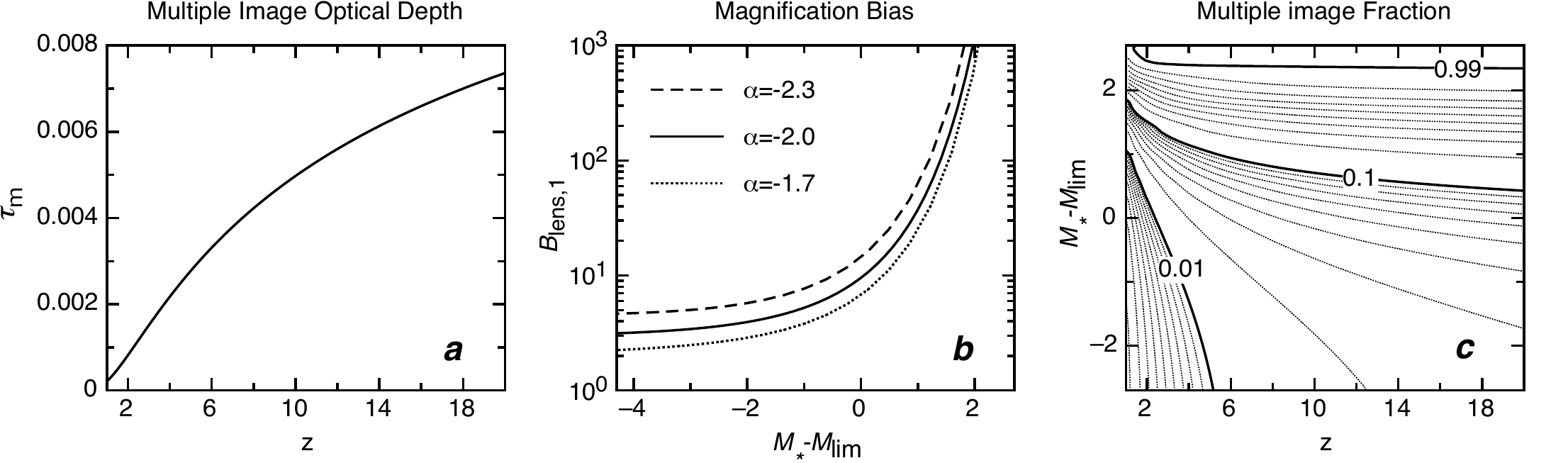}
\end{center}
\vspace{-1mm}
Supplementary Figure 2. {\bf Probabilities for multiple imaging of high redshift
galaxies.} {\bf Panel a:} The lensing optical depth as a function of
redshift. {\bf Panel b:} The magnification bias as a function of the difference
between $M_\star$ and the limiting survey absolute magnitude $M_{\rm lim}$. 
Three values of the faint-end LF-slope $\alpha$ are considered. {\bf Panel c:}
Contours of $F_{\rm lens}$ as a function of $z$ and ($M_\star-M_{\rm lim}$),
assuming$^1$ $\alpha=-2$.  
\end{figure*}

\begin{figure*}
\vspace{10mm}
\begin{center}
\includegraphics[width=17cm]{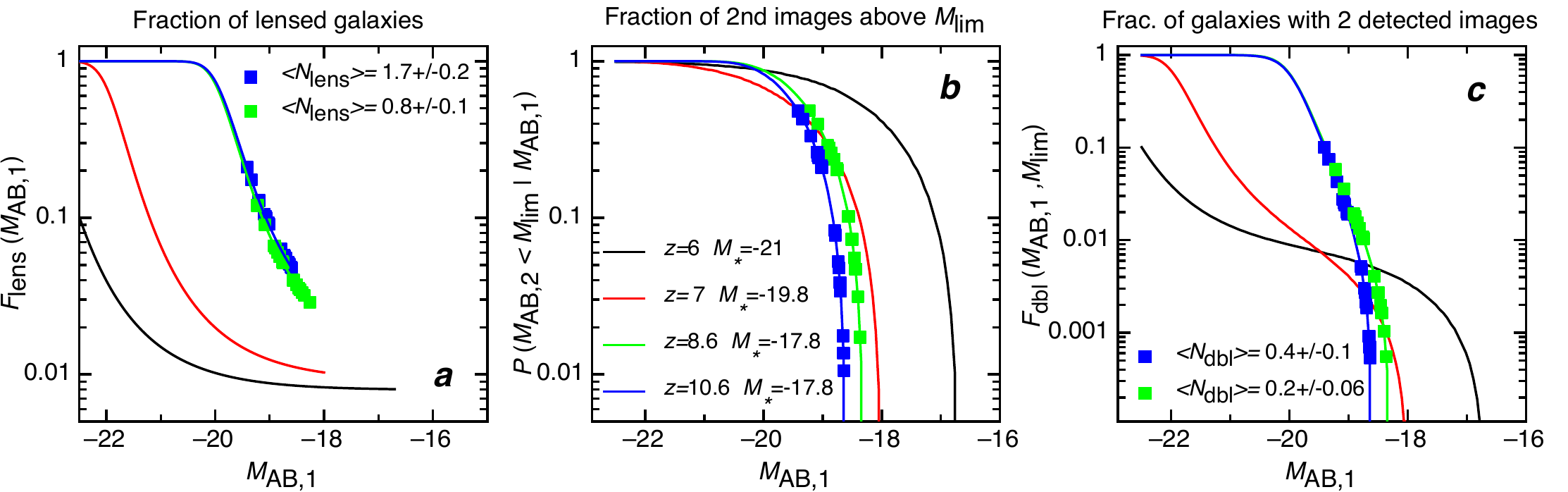}
\end{center}
\vspace{-1mm}
Supplementary Figure 3. {\bf Probabilities for multiple imaging of $z\simeq8-10$ galaxy candidates.} {\bf Panel a:} The probability that a galaxy with observed
$M_{AB,1}$ is multiply-imaged. The expected mean number of lenses $(\langle N_{\rm lens}\rangle $) among the $z\simeq8.6$ and $z\simeq10.6$ candidates is listed. {\bf Panel b:} The probability that a {\em
lensed} galaxy with observed $M_{AB,1}$ has a corresponding second image with
$M_{AB,2}<M_{\rm lim}$. {\bf Panel c:} The fraction of galaxies that are part
of a lensed pair in which both images are detectable (68\% errors here were
computed using a bootstrap method). The expected mean number of systems that would be observed as doubles ($\langle N_{\rm dbl}\rangle$) is listed. We have assumed the
determinations of $M_\star=-17.8$ and $\alpha=-2$, and observed absolute magnitudes $M_{AB,1}$ from Yan et al.$^{4}$ (the
squares correspond to probabilities for the individual galaxy candidates). 
\end{figure*}

\begin{figure*}
\begin{center}
\vspace{-5mm}
\includegraphics[angle=0,height=10.3cm]{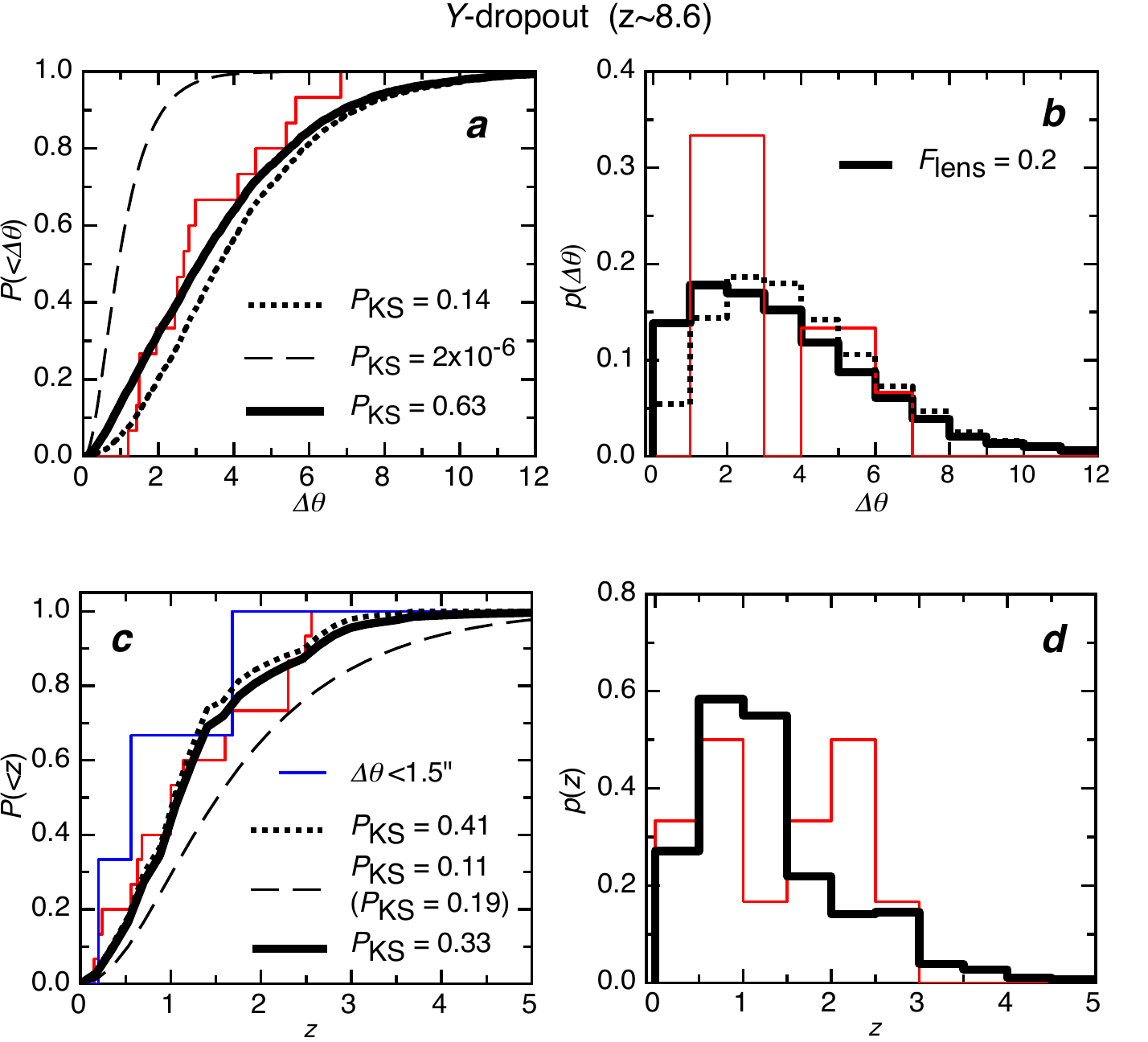}

\end{center}
Supplementary Figure 4. {\bf
Probability distributions for angular proximity and redshift of bright
foreground galaxies  among the sample of $z\sim8.6$ candidates.} 
{\bf Panel a:} The cumulative distribution for the angular separation between $z\simeq8.6$ candidates and
their nearest foreground galaxies with $H\leq25$ in the {\em HUDF} (red histogram). Also shown are the model cumulative distributions of
angular separations between {\it random} lines-of-sight and the nearest bright
foreground galaxies (dotted black line), and of angular separations for
the brighter image of gravitationally {\it lensed} objects at $z=8.6$ (dashed 
black  line). The thick black line shows the composite cumulative distribution
generated by summing the {\it random} and {\it lensed} histograms, with a
weight equal to a lens fraction of $F_{\rm lens}=0.2$. {\bf Panel b:} The binned histograms (area normalised to unity) for the
angular separations of observed candidates (red), for separations in the composite model (thick black), and for separations from random lines of sight (dotted black). 
 {\bf Panel c:} The cumulative redshift distribution for foreground
galaxies associated with $z\simeq8.6$ candidates (red histogram). Also shown are the cumulative distributions for the
redshifts of foreground galaxies nearest to {\it random} lines of sight (dotted black line),
and for the expected gravitational lens redshifts assuming sources at $z=8.6$
(dashed black line). The thick black line shows the composite cumulative
distribution ($F_{\rm lens}=0.2$). We
also plot the redshift distribution of foreground galaxies within 1.5
arcseconds of a $z\simeq8.6$ candidate (blue histogram). {\bf Panel d:} The binned
histograms for the foreground galaxy redshifts along lines of sight to dropout candidates
 (red), and for the composite model (black).  In
each case, values of $P_{\rm KS}$ corresponding to the comparison of the data
with the model distributions are listed.
\end{figure*}

\begin{figure*}
\begin{center}
\includegraphics[angle=0,height=10.3cm]{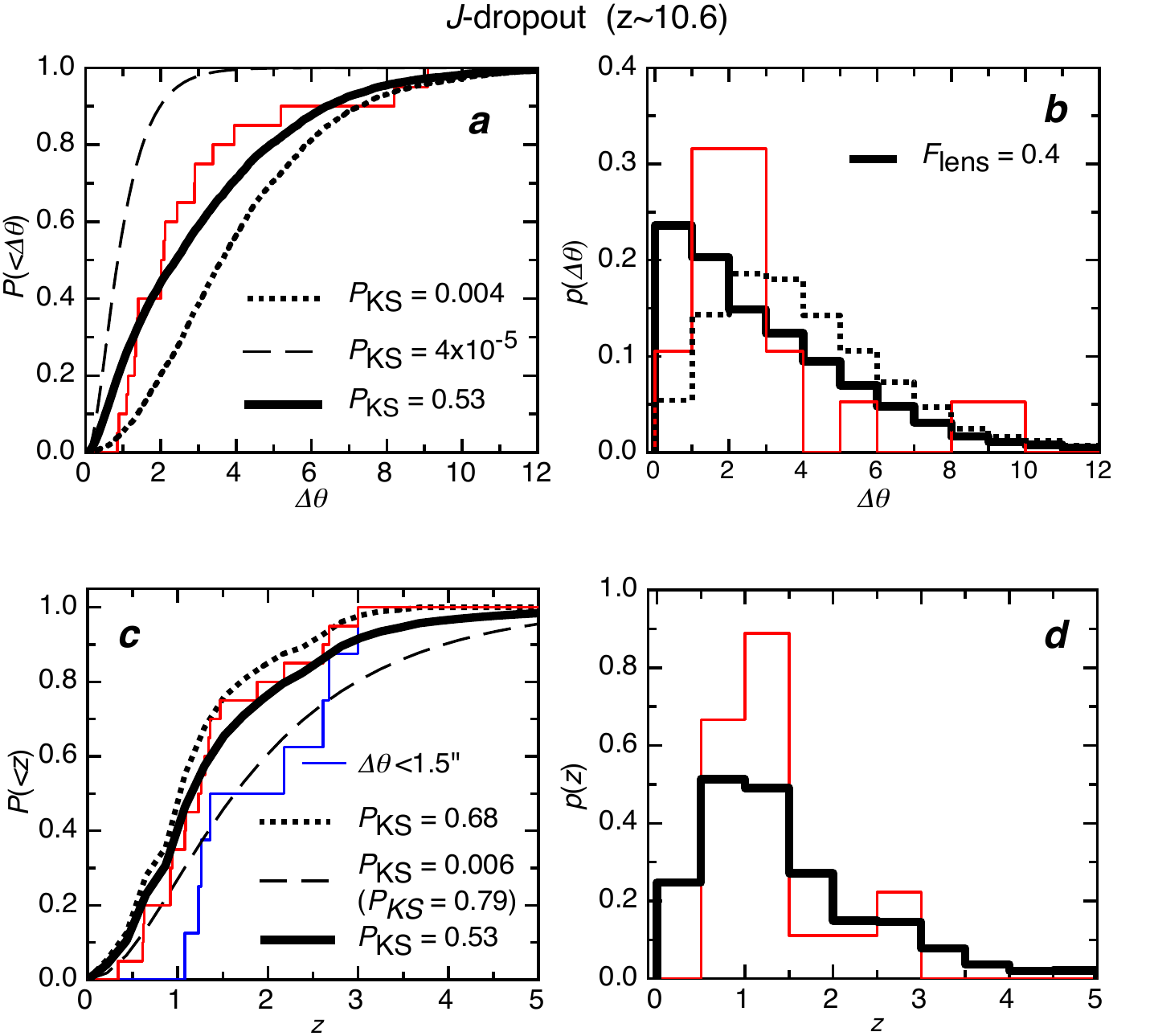}
\end{center}
\vspace{-5mm}
Supplementary Figure 5. {\bf
Probability distributions for angular proximity and redshift of bright
foreground galaxies among the sample of $z\sim10.6$ candidates.} The panels mirror those of Supplementary Figure~4. We assume $F_{\rm lens}=0.4$ for the model composite distribution. 
\end{figure*}

\end{document}